\documentclass[12pt,dvipsnames]{iopart}

 \expandafter\let\csname equation*\endcsname\relax 
 \expandafter\let\csname endequation*\endcsname\relax 
\usepackage{amsmath,amsfonts,amssymb}

\usepackage{graphicx,subfloat}
\usepackage{color}
\usepackage{comment}
\usepackage{appendix}
\usepackage{multirow}
\usepackage{epstopdf}
\usepackage{etoolbox}
\usepackage{hyperref}
\usepackage{grffile}

\makeatletter
\renewcommand\@appendixstar{\@@par
 \ifnumbysec 
 \@addtoreset{table}{section}
 \@addtoreset{figure}{section}\fi
 \setcounter{section}{0}
 \setcounter{subsection}{0}
 \setcounter{subsubsection}{0}
 \setcounter{equation}{0}
 \setcounter{figure}{0}
 \setcounter{table}{0}
 \def\thesection{Appendix \Alph{section}} % this line has been \def\thesection{Appendix \Alph{section}} before
 \def\theequation{\ifnumbysec
      \Alph{section}.\arabic{equation}\else
      \Alph{section}\arabic{equation}\fi}
 \def\thetable{\ifnumbysec
      \Alph{section}\arabic{table}\else
      A\arabic{table}\fi}
 \def\thefigure{\ifnumbysec
      \Alph{section}\arabic{figure}\else
      A\arabic{figure}\fi}}
\makeatother
\newcommand{\bb}{{\bf b}}

\newcommand{\bk}{{\bf k}}
\newcommand{\bl}{{\bf l}}
\newcommand{\bm}{{\bf m}}

\newcommand{\bv}{{\bf v}}

\renewcommand{\d}{\text{d}}

\newcommand{\mE}{\mathcal E}
\newcommand{\mF}{\mathcal F}

\newcommand{\mH}{\mathcal H}
\newcommand{\mL}{\mathcal L}

\newcommand{\mR}{\mathcal R}

\newcommand{\resp}{\emph{resp.} }

\newcommand{\av}[1]{\langle {#1}\rangle}

\newcommand{\ave}[1]{\langle {#1}\rangle}
\newcommand{\avt}[1]{\langle {#1}\rangle_{\tilde \rho }}

\newcommand{\np}{{n^\prime}}
\renewcommand{\lo}{{l_0}}

\newcommand{\atanh}{{\text{atanh }}}

\newcommand{\modif}[1]{{#1}}
\newcommand{\citeB}[1]{(see \cite{#1})}

\begin{document}
\graphicspath{}
\title{Optimal thermalization in a shell model of homogeneous turbulence.}
\date{\today}
\author{Simon Thalabard  and Bruce Turkington} 

\address{Department of Mathematics and Statistics, University of Massachusetts, Amherst, MA 01003, USA.}
%\address
\date{\today}

\begin{abstract}

We investigate the  turbulence-induced dissipation of the large scales in a statistically homogeneous flow using an 
``optimal closure,'' which  one of us (BT) has  recently exposed in the context of Hamiltonian dynamics.     
This statistical closure employs a Gaussian model for the turbulent scales, with corresponding vanishing third cumulant, 
and yet it captures an intrinsic damping. The key to this apparent paradox lies in a clear distinction between true ensemble averages and their proxies, most easily grasped when one works directly with the Liouville equation rather than the cumulant hierarchy. 
%Schematically said, closure is achieved by minimizing (in an appropriate sense)  a lack-of-fit residual, which retains the intrinsic features  of the true dynamics.
We focus on a simple problem for which the optimal closure can be fully and exactly worked out: 
the relaxation arbitrarily  far-from-equilibrium of a single energy shell towards  Gibbs equilibrium 
in an inviscid shell model of 3D turbulence.     The predictions of the optimal closure are validated against 
DNS and contrasted with those derived from EDQNM closure.   
%We then extend our calculation to the full 3D Galerkin-Euler dynamics.
\end{abstract}
%\large
\maketitle
\newcommand{\eqspace}{\hspace{0.3cm}}

\section{Introduction}
The concept of a turbulent cascade -- be it direct or inverse -- is tied to non-Gaussianity.  In the case of 3D homogeneous isotropic turbulence, this is patent from Kolmogorov 4/5 law, which states that the rate of energy dissipation  is proportional to the skewness  of the velocity increment.  Be the deviations from Gaussianity weak (as in the 2D inverse cascade of energy) or strong (as in the 3D direct cascade),  Gaussian models of turbulence are therefore paradoxical, in that they cannot fully represent turbulent fluctuations
and interactions.
The paradox is most evident in the so-called  Quasi-Normal (QN) cumulant discard closure proposed by \cite{millionshtchikov_role_1941}, which relies on a Gaussian Ansatz to close the cumulant hierarchy.   There, in order  to provide a non-trivial closed equation  for the energy spectrum, Gaussianity needs to be used inconsistently, that is, for the fourth cumulant only and not for the third.
Still, the QN theory is non-realizable, as it leads to the development of non-spurious negative energy spectra   \citeB{kraichnan_relation_1957,orszag_lectures_1977}.
Orszag showed that the QN theory can be patched into the now-celebrated Eddy Damped Quasi Normal Markov theory (EDQNM), by taking the Markovian limit and introducing phenomenological eddy damping.    
While EDQNM provides a practical theory that has been widely used in the past decades for the purpose of subgrid modeling \citeB{lesieur_turbulence_2008,baerenzung_spectral_2008}, it remains conceptually unsatisfactory as it relies on an ad-hoc modeling of the statistics to correct the defects of the QN theory. %
\modif{Besides, EDQNM does not completely resolve the realizability issue : counter-examples can be constructed  in dimension $d \le 2$, where a negative energy spectrum develops for arbitrarily small times \citeB{fournier_$d$-dimensional_1978}}.

The reason why the QN theory fails so badly remains unclear.  Is it due to the Gaussian Ansatz or rather its inconsistent use~?
More generally, is the choice of a Gaussian as a proxy for the true ensemble really bound to yield either a trivial or an ad-hoc closure~?   
The purpose of this paper is to answer this question in the negative.   To this end, we rely on a unconventional  
``optimal closure framework'' recently proposed by one of us in the context of Hamiltonian systems (\cite{turkington_optimization_2013}).   When applied to turbulence problems, it allows a consistent use of a Gaussian Ansatz 
(and corresponding vanishing third cumulant) without loss of the intrinsic dissipation due to turbulence.  
This apparent paradox is resolved by making a clear distinction between the true ensembles and their proxies. 
While this distinction is not transparent in the cumulant hierarchy, it is revealed when one works directly with the Liouville
equation.  
Schematically said, any attempt to model the unresolved turbulent scales comes with a cost, which can be quantified via
a residual in the Liouville equation.   The idea of the optimal closure is  to make this cost minimal in an appropriate sense. 

In this paper, the exposition of this closure will be practical rather than general.  While there exist intimate connections between the optimal closure and general out-of-equilibrium frameworks, those will not be discussed here, and we refer the  reader to \cite{turkington_optimization_2013} for a thorough exposition.   For the sake of clarity, we study a simple test problem,
for which closure can be entirely implemented, namely, 
the relaxation arbitrarily far from equilibrium of an energy shell towards Gibbs equilibrium in a Galerkin-truncated, inviscid,   statistically homogeneous fluid.  
The one-mode relaxation has been considered by \cite{krstulovic_two-fluid_2008} for Galerkin-truncated 3D dynamics. 
They showed that the relaxation could be understood in terms of a two-fluid model, where thermalization is caused by the small-scale dynamics, and that insightful predictions could be obtained by EDQNM. 
To emphasize the key issues in the closure theory, without obscuring them in the mathematical analysis,  here we focus on a shell model dynamics, which mimics the nonlinearity of the 3D truncated Euler dynamics.  In this specific problem, not only does the optimal strategy yield a consistent closure, it also predicts explicitly the relaxation profile and the time-scales involved.

 From a physical point of view,  inviscid Galerkin-truncated dynamics have been invoked to describe the interactions between thermalized and non-thermalized degrees of freedom.  In this sense such systems are now known to provide minimal models of turbulence, with the highest wavenumbers typically acting as a bath of energy \citeB{cichowlas_effective_2005,bos_dynamics_2006,krstulovic_cascades_2009}.   This feature was in fact early noticed by  Kraichnan \citeB{kraichnan_is_1989,eyink_robert_2010}, and lies at the heart of  his  Fluctuation-Dissipation theorem \citeB{kraichnan_classical_1959}, later exploited by  Leith for the purpose of climate modeling \citeB{leith_climate_1975,bell_climate_1980}.  Truncated dynamics also appear in the modeling of quantum turbulence, where bottlenecks of energy between quantum and classical degrees of freedom yield a partial thermalization of an intermediate range of scales \citeB{di_molfetta_self-truncation_2015}.
 \modif{
While the long-time properties of truncated dynamics can be described by standard equilibrium statistical mechanics, the non-equilibrium transients are far less understood, as they involve the interplay between conservative and dissipative dynamics \cite{frisch_hyperviscosity_2008,banerjee_transition_2014}.
In recent years, the increase of computer of power and systematic use of direct numerical simulations has reinvigorated the subject and motivated new research directions. They have been used to probe the relevance of renormalization group techniques and associated perturbation expansions \cite{frisch_turbulence_2012,lanotte_turbulence_2015}, and they have also revealed intriguing decaying phenomenologies~: from ``Tyger-resonances'' that pre-curse thermalization \cite{ray_resonance_2011,ray_thermalized_2015} to  anomalous spectral  laws found in so-called Leith models of turbulence \cite{connaughton_warm_2004, thalabard_anomalous_2015}.}
  
The paper is organized as follows. We first recall a few salient points about the shell dynamics that we use, and expose the general out-of-equilibrium set-up that we consider. Second, we outline the optimal closure procedure, and solve the one-mode relaxation problem. We then contrast its predictions with those obtained from the EDQNM framework. %
The skill of the optimal closure is tested against numerical simulations.   We end by briefly suggesting how this approach could be extended to more complex problems. 

\section{Test problem setup}
\subsection{Shell dynamics}
In order to keep the algebra involved in the derivation of the optimal closure as simple as possible, we study a class of 
``shell dynamics'', which slightly extends the popular GOY dynamics initially proposed by \cite{gledzer_system_1973,yamada_lyapunov_1988}.   This model is known to reproduce many of the statistical signatures of high-reynolds turbulent cascades (see \cite{pisarenko_further_1993,biferale_shell_2003}), among which is intermittency.  
The  dynamical  variables $v_l$ ($1\le l \le N$) represent the typical velocity for the wavelength $k(l)$. They  are taken to obey a very general nonlinear quadratic dynamics
\begin{equation}
 \dot v_l = C_{lmn} \gamma_n v_m^\star v_n^\star \underset{Def}{=} \mF_l[v].
 \label{eq:CompactShell}
\end{equation}
Here and later, stars denote complex conjugation, and  summations over pairs of dummy indices are implied.  
The coefficients $\gamma_n$ and $C_{lmn}$ are yet to be prescribed.  
We first require that the nonlinear dynamics  (\ref{eq:CompactShell}) preserves the  two following quadratic invariants~:
\begin{equation}
\mE[v]=\sum_{l=1}^N |v_l |^2 \text{ and } \mH[v]=\sum_{l=1}^N \gamma_l |v_l|^2.
\label{eq:EHDef}
\end{equation}
To  make $\mE$ and $\mH$ broad analogues of the kinetic energy and the helicity, preserved by the 3D Euler dynamics, 
dimensional analysis suggests setting 
  $\gamma_l = (-1)^l k(l)$. 
Independently of this choice, the conservation of both the energy and the helicity is achieved by requiring the coefficients 
$C_{lmn}$ to be fully anti-symmetric, namely, 
\begin{equation}
C_{lmn} +C_{lnm} = 0 \text{~and~} C_{lmn}=C_{mnl}=C_{nlm} \text{~for any $l,m,n$.}
\label{eq:ShellCoeffproperties}
\end{equation} 
To  specify dynamics completely, we now make the standard assumption that  the triadic interactions between the shells are local: the variables $v_l$ ($1\le l \le N$) are nonlinearly coupled to their nearest and next-nearest neighboring shells only.   The coefficients $C_{lmn}$ are then zero except when $\left \lbrace l,m,n \right \rbrace$ is a set of consecutive integers. The remaining  coefficients are then entirely determined by the value of the coefficients $C_{(l-1)l(l+1)}$, using the anti-symmetry properties (\ref{eq:ShellCoeffproperties}).  We choose  
\begin{equation}
C_{(l-1)l(l+1)} = \dfrac{k (l+1)}{\gamma_{l}-\gamma_{l-1}} \;\;\; \text{~if~} 1<l < N, \hspace{0.5cm}\text{and~} 0 \text{~otherwise.}
\label{eq:ShellCoeffs}
\end{equation}
By further choosing logarithmically spaced shells  $k(l)=k_0 \lambda^l$, one retrieves exactly the standard formulation of GOY model found for example in \cite{biferale_shell_2003}; that is, up to a global rescaling $v \to -i v /\lambda^3$ that we  omit here.

\subsection{Relaxation Towards Equilibrium}
\begin{figure}
\begin{minipage}{0.4\textwidth}
\centering
\begin{center}
\hfill $t=0$ \\
\vspace{0.3cm}
\includegraphics[width=\textwidth]{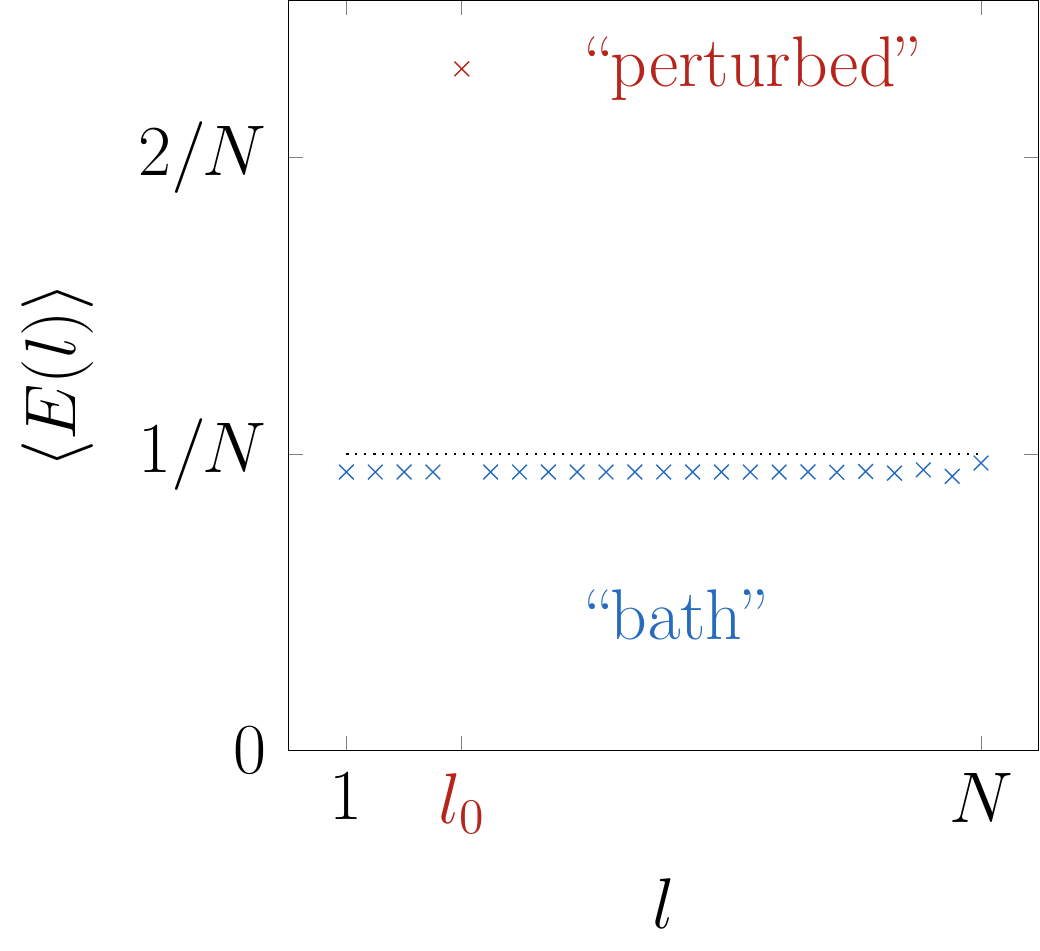}
\end{center}
\end{minipage}
\begin{minipage}{0.2\textwidth}
\centering
\Huge $\longrightarrow$
\end{minipage}
\begin{minipage}{0.4\textwidth}
\centering
\hfill  $t=\infty$ \\
\vspace{0.3cm}
\includegraphics[width=\textwidth]{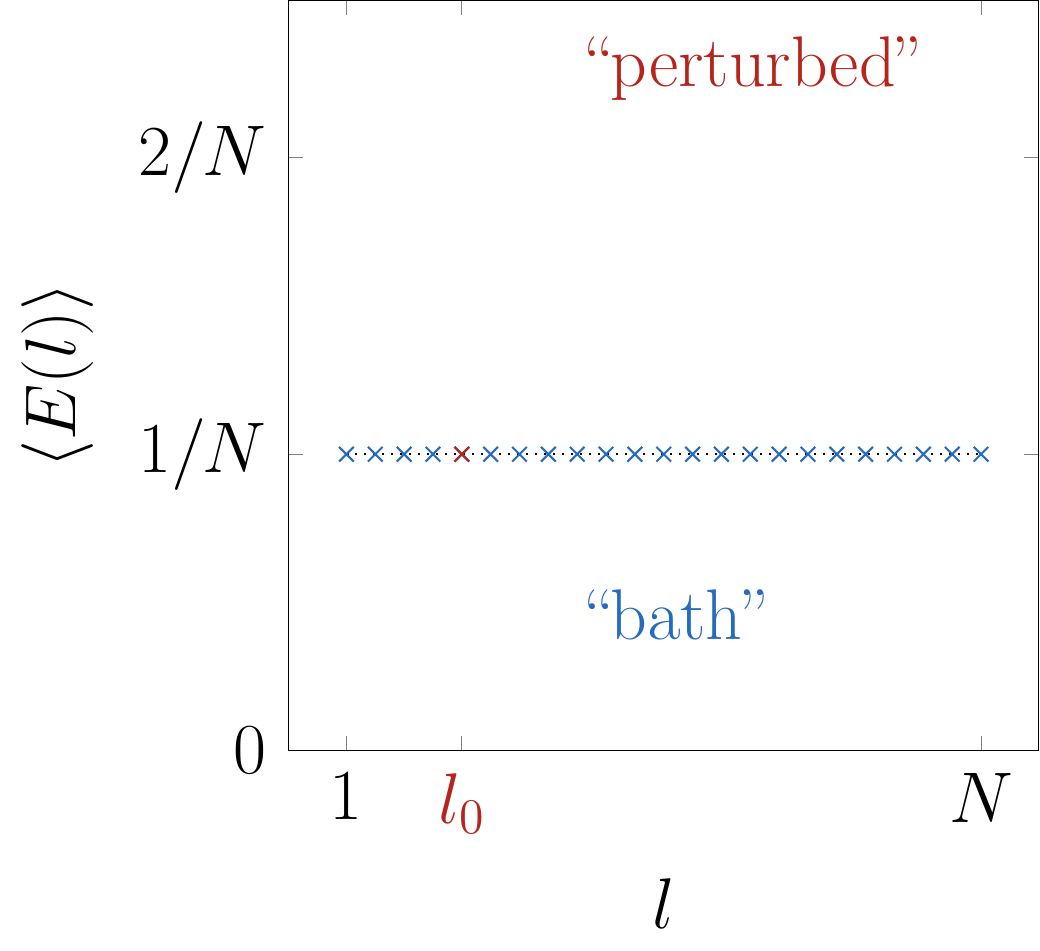}
\end{minipage}
\caption{Cartoon of the out-of-equilibrium setup, with total energy $\ave{\mE}=1$. At time $t=0$, the mode $l_0$ is set away  from equipartition:  $\ave{\mE[l_0]} \gg 1/N$ . The excess energy is distributed among the modes $l \neq l_0$, which constitute the ``bath'' of energy.  The averaged helicity is  $\ave{\mH} = H_{equi} $ (see definition in text).}
\label{fig:setup}
\end{figure}

In addition to conserving the two quadratic invariants (\ref{eq:EHDef}), the dynamics (\ref{eq:CompactShell}) trivially satisfies  a detailed Liouville theorem. In particular, this implies  that  Gibbs ensembles are statistical invariants for the dynamics,  
the Gibbs distribution for  (\ref{eq:CompactShell}) being defined in terms of the ``inverse temperatures" $\beta_l$ as  
\begin{equation}
\begin{split}
& \rho_G = \dfrac{1}{Z} e^{-\beta E -\zeta H}=\prod_{1\le l \le N} \rho_{G,l},  \\ 
&\text{~with~} \;\; \rho_{G,l} = \dfrac{\beta_l}{\pi} e^{-\beta_l |v_l|^2}, \text{~and~}  \beta_l = \beta + \zeta \gamma_l.
\end{split}
\label{eq:gibbs}
\end{equation}
In principle,  we expect that any non-degenerate ensemble of solutions with  prescribed ensemble-averaged  energies and helicities converges towards the corresponding Gibbs ensemble in a statistical sense, no matter how far  the initial ensemble is from equilibrium.  
In particular,  we focus our attention on  nonequilibrium initial ensembles 
  generated at time $t=0^+$  by taking  the inverse temperature of one specific \emph{perturbed} mode $\lo$  far away from its equilibrium Gibbs value. In this set-up, the $ N-1$ other modes can be thought of as a \emph{bath} of energy and helicity. The total energy (\resp the total helicity) is then the sum of two terms :  The perturbed energy $\av{\mE_{\lo}}$  (\resp  the perturbed helicity 
  $\av{\mH_{\lo}}$) and the bath energy $\av{\mE_{B}}$ (\resp the bath helicity $\av{\mH_{B}}$)).
For example, we may choose to make the system relax towards the \emph{equipartition} ensemble, corresponding to the case 
$\zeta=0$ in (\ref{eq:gibbs}) and corresponding total helicity  $H_{equi} = -E \left[\lambda+(-\lambda)^{N+1} \right] \left[1+\lambda \right]^{-1} $. Setting $\av{\mH_B}=H_{equi}-\av{\mH_\lo}$ and  $\av{\mE_B}=1-\av{\mE_\lo}$ at initial time, we then expect the system to relax towards the equipartition ensemble, a situation  cartooned in Figure \ref{fig:setup}.
This is precisely the relaxation process that we plan to  approximate using a closed and tractable set of equations.
\section{The optimal closure framework}
\subsection{General philosophy}
Fully and accurately  describing the relaxation of the probability density $\rho(t,\cdot)$ is clearly out of reach. 
It would require solving the following Liouville equation,
\begin{equation}
 \left[ \partial_t + \mL \right] \log \rho(t,\cdot) =0  \text{~~with~~} \mL =\mF_l \partial_{z_l}+ \mF_l^\star \partial^\star_{z_l^\star},
 \label{eq:LiouvilleEq}
\end{equation}
a task that involves as much (if not more) analytical and computational effort as solving the original dynamics.
A more accessible path is to look for approximate solutions of the Liouville equation, a strategy  that does not require solving the entire dynamics, but instead relies on a closure hypothesis. More specifically, we make the following seemingly naive Ansatz, that  throughout the relaxation the probability densities can be approximated by time-dependent Gaussians, 
\begin{equation}
\tilde \rho(\bb(t),\cdot) =\prod_{1\le l \le N} \tilde \rho_{l}(b_l(t),\cdot ), \;\; \text{~with~}  \;\; \tilde \rho_{l}(b_l(t),\cdot) = \dfrac{b_l(t)}{\pi} e^{-b_l(t) |v_l|^2}, 
\label{eq:Ansatz}
\end{equation}
that are entirely determined by the inverse temperature vector $\bb = (b_1,...,b_N)$.
We will write $\avt{\cdot}$ for averages with respect to the distribution $\tilde \rho$. 

It is obvious, yet  crucial, to note that  those tilded averages \emph{do not} in general match the true ensemble averages $\av{\cdot}$, that is, averages with respect to the density $\rho$. While valid for the specific times $t=0^+$ and $t \to +\infty$,  a crude identification between $\av{\cdot}$ and  $\avt{\cdot}$ is certainly not justified  for  intermediate times. This is patent from the equation for the energy shell evolution, \emph{viz.},  
 \begin{equation}
\dfrac{\d}{\d t�} \av{|v_l|^2}  =2 C_{lmn} \gamma_n \Re \left[  \av{v_l^\star  v_m^\star  v_n^\star } \right],
\label{eq:secondmoment}
\end{equation}
whose r.h.s would  then vanish, preventing any relaxation!  The essential observation is that the densities $\tilde \rho$  do not 
necessarily satisfy the Liouville equation  (\ref{eq:LiouvilleEq}), and that their error can be quantified in terms of the \emph{Liouville Residual}: 
 \begin{equation}
\mR[\bb(t),\cdot] =  \left[ \partial_t + \mL \right] \log \tilde \rho \neq 0 .
 \label{eq:LiouvilleRes}
\end{equation} 
This statistical quantity  can be interpreted as the local rate of information loss for proxying $\rho$ by $\tilde \rho$ ( \cite{turkington_optimization_2013}).
Therefore, using  $\tilde \rho$ as a proxy for $\rho$  does not necessarily make the r.h.s of \label{eq:secondmoment} disappear, the averages being  formally related through
\begin{equation}
\av{\cdot}  = \avt{e^{-\int_0^t e^{-\tau \mL}\mR(t-\tau)\d \tau} \cdot } \neq \avt{\cdot}.
\end{equation}

Quite generally, the Liouville Residual can be written as the sum of two contributions, which we label  $\mR_r$ and $\mR_i$. The subscripts stand for ``\emph{r}eversible'' and ``\emph{i}rreversible''. For the dynamics (\ref{eq:CompactShell}), the Liouville residual associated to the Gaussian Ansatz (\ref{eq:Ansatz}) is readily obtained from the definition  (\ref{eq:LiouvilleRes}) and a few lines of algebra. It reads 
\begin{equation}
\mR(\bb,\dot \bb,\bv) =  \underbrace{\dot b_l \left( \dfrac{1}{b_l}-|v_l|^2  \right)}_{\mR_r} + \underbrace{ (-2) b_l \Re \left[ C_{lmn} \gamma_n  v_l^\star v_m^\star v_n^\star \right]}_{\mR_i}.
\label{eq:ShellProj}
\end{equation}

The philosophy of the optimal closure is to pick the inverse temperature profile $\bb(t)$  whose dynamics provide a closest match to the actual Liouville dynamics.  That is,  the optimal profile minimizes the rate of information loss over the relaxation.
To achieve this, we associate to each Ansatz a  \emph{lack-of-fit}  Lagrangian density, which we define up to a weighing constant $\kappa^2 > 0$ as 
\begin{equation}
L_{lof}[\bb,\dot \bb] = \dfrac{1}{2} \avt{\left[ \mR_r + \kappa \mR_i \right]^2} = \underbrace{ \dfrac{1}{2} \avt{\mR_r^2}}_{L_r} + \underbrace{ \dfrac{\kappa^2}{2} \avt{\mR_i^2}}_{L_i}.
\label{eq:defLOF}
\end{equation}
The closure is then defined by solving the optimization problem:
\begin{equation}
 \underset {\bb(t)} 
 \min \int_{0}^\infty \d t \,  {L}_{lof} [\bb,\dot \bb] \hspace{0.3cm}\text{~over paths  $\bb(t)$ subject to~}  
 \begin{cases}
 &b_l(0^+)=b_l^{0} \\
 & b_l(\infty)=  \beta_l
 \end{cases}
 %\text{for any } 1\le l\le N%
 .
 \label{eq:MIN}
\end{equation}
The lack-of-fit Lagrangian density is expressed as the sum of a ``reversible'' density $L_r$ with the ``irreversible''  density $L_i$.
By analogy to standard Hamiltonian mechanics, 
the reversible contribution is akin to a kinetic energy, while the irreversible contribution may be thought of minus the potential energy.    The nonequilibrium thermodynamics  associated with this optimization formulation 
is discussed by \cite{turkington_optimization_2013} and \cite{kleeman_path_2015}. 

\subsection{The quasi-normal optimal closure.}
\subsubsection{The lack-of-fit Lagrangian}
We use the general expression (\ref{eq:ShellProj})  for the Liouville Residual, together with the Gaussian Ansatz (\ref{eq:Ansatz}) to obtain the lack-of-fit Lagrangian. Some lines of algebra then yield specifically :
 \begin{equation}
 \begin{split}
& L_r=\dfrac{1}{2} \dfrac{\dot b_l^2}{b_l^2} 
 \;\; \text{~and~}  \;\; L_i=\dfrac{\kappa^2}{2} \left\lbrace  \dfrac{g(l)}{b_l} + f(l,m,n)\dfrac{b_l}{b_m b_n} \right \rbrace\\
\end{split}.
\label{eq:ShellLagrangian}
 \end{equation}
for  functions $f$ and $g$ that  are found to be 
\begin{equation}
 f(l,m,n)= C^2_{lmn} \gamma_n(\gamma_n-\gamma_m) \;\; \text{~ and~} \;\; g(l)={C^2_{lmn}}\left\lbrace 2 \gamma_m\gamma_l-\gamma_m\gamma_n-\gamma_l^2 \right\rbrace.
\label{eq:fg}
\end{equation}
\subsubsection{Single mode relaxation towards equipartition}
We now consider the specific scenario cartooned in Figure \ref{fig:setup}, where a single mode $\lo$ relaxes towards the  equipartition value $\beta = N/E$, obtained by setting $\zeta = 0 $ in the definition  (\ref{eq:gibbs}) of Gibbsian ensembles.   We  further assume that the Lagrangian depends only on  two temperatures: the temperature associated to the perturbed mode $1/b_{\lo}$ and the bath temperature $1/b_B$, so that  
\begin{equation}
 b_l(t) = b_B(t) \text{~~for all~~} l \neq \lo.
\label{eq:BathTemp}
\end{equation}
For this single-mode relaxation we can entirely determine the temperature profile that solves the optimization problem  (\ref{eq:MIN}). 
From Equations (\ref{eq:BathTemp}) and (\ref{eq:ShellLagrangian}), we find that the Lagrangian density
can be recast in terms of a single characteristic parameter $\phi_\kappa(l_0)$ as follows: 
\begin{equation}
\begin{split}
& L_r =\dfrac{1}{2} \dfrac{\dot b_\lo^2}{b_\lo^2} + \dfrac{N-1}{2} \dfrac{\dot b_B^2 }{b_B^2} \;\; \text{~and~}  \;\; 
 L_i  =\dfrac{\phi_\kappa(\lo)}{2} \left\lbrace  - \dfrac{2}{b_B} +  \dfrac{1}{b_\lo} +
\dfrac{b_\lo}{b_B^2} \right \rbrace, \\
& \text{~with ~}  \;  \phi_\kappa(\lo)  = \kappa^2 \sum_{1 \le m,n \le N}f(\lo,m,n) .
\end{split}
\label{eq:SimplifiedIrr}
\end{equation}
The solution to the optimization problem  (\ref{eq:MIN}) is obtained from the Euler-Lagrange equations associated to the Lagrangian density (\ref{eq:SimplifiedIrr}). 
To this end, it is easier to work with the  following thermodynamic variables:
\begin{equation} 
  \sigma_\lo = \log \dfrac{b_\lo}{\beta} \text{~and ~} \sigma_B= \sqrt{N-1} \log \dfrac{b_B}{\beta},
  \label{eq:SigmaVar}
 \end{equation}
so that the Lagrangian density now reads
\begin{equation}
 L_{lof}= \dfrac{\dot \sigma_\lo^2}{2} + \dfrac{\dot \sigma_B^2}{2} + \dfrac{\phi_{\kappa}(\lo)E}{2 N}\left\lbrace  -2 e^{-\frac{\sigma_B}{\sqrt{N-1}}} +  e^{-\sigma_\lo}+ e^{\sigma_\lo-2\frac{\sigma_B}{\sqrt{N-1}}} \right\rbrace.
\end{equation}
Assuming $N \gg 1$,  we obtain the temperature dynamics to first order in $1/N$ as  
\begin{equation}
\dot \sigma_B=0 \;\; \text{~and~}  \;\; \dot \sigma_\lo = -2 \sqrt{ \dfrac{\phi_\kappa(\lo)E}{N}} \sinh \dfrac{\sigma_\lo}{2}.
\label{eq:Sigrelax}
\end{equation}
The relaxation profile is then given by 
\begin{equation}
 \tanh \dfrac{\sigma_\lo}{4} = \tanh \dfrac{\sigma_\lo(0^+)}{4} e^{ - 2 t/\tau_\kappa(\lo)} \; \; \text{~with~}  \; 
 \tau_\kappa(\lo)=\sqrt{\dfrac{4N}{\phi_\kappa(\lo)E}}.
 \label{eq:relax}
\end{equation}
In terms of the inverse temperatures, 
\begin{equation}
b_\lo (t)= \beta  \left\lbrace  \dfrac{\sqrt{b_\lo^0/\beta} + \tanh  t/\tau_\kappa(\lo) }{\sqrt{b_\lo^{0}/\beta} \tanh t/\tau_\kappa(\lo)+ 1}\right\rbrace^{2}.
\label{eq:relaxTemp}
\end{equation}
%
%\begin{equation}
%\begin{split}
% & \tanh \dfrac{\sigma_0}{4} = \tanh \dfrac{\sigma_0(0^+)}{4} e^{ - 2 t/\tau_\kappa(\lo)}, \\
%  & \text{ie, } b_\lo (t)= \beta  \left\lbrace  \dfrac{\sqrt{b_\lo^0/\beta} + \tanh  t/\tau_\kappa(\lo) }{\sqrt{b_\lo^{0}/\beta} \tanh t/\tau_\kappa(\lo)+ 1}\right\rbrace^{2}, \text{~with~} \tau_\kappa(\lo)=\sqrt{\dfrac{4N}{\phi_\kappa(\lo)E}}.
%  & \text{ie, } b_\lo (t)= \beta \tanh^2 \left[ t+T_0\right], \text{~with~} \tau[\lo]=\sqrt{\dfrac{4N}{\phi[\lo]E}}.
%\end{split}
%\label{eq:relax}
%\end{equation}
The  optimal closure therefore predicts a  $\tanh$-decay of the energy perturbation  $E_\lo(t) = 1/b_\lo(t)$, namely :  
\begin{equation}
\begin{split}
E_\lo(t)=(1/\beta) \tanh^{\pm 2 }&\left( \Delta_0 + t/\tau_\kappa(\lo)\right) \text{~~with~~} \Delta_0 = \atanh\left[ \left(b_\lo^0/\beta \right)^{\mp1/2}\right], \\
&\text{and~} \;\;  \pm = \text{sign}  \left(b_\lo^0/\beta-1 \right).
\end{split}
\end{equation}
We finally note that for the GOY dynamics, the constants $\phi_\kappa(l)$ can be explicitly determined from Equations (\ref{eq:ShellCoeffs})  and  (\ref{eq:SimplifiedIrr}) as 
\begin{equation}
\phi_\kappa(l) = \kappa^2 k_0^2 \lambda^{2l} \cdot \left[\lambda^6 + \lambda^2(\lambda-1)^2 +1\right].
\label{eq:relaxCste}
\end{equation}
\subsection{Turbulent dissipation and asymptotics}
The set of predictions (\ref{eq:Sigrelax})-(\ref{eq:relaxCste}) is not trivial. In particular, it is worth emphasizing that the $\tanh$ ``optimal decay''  is  very different from standard exponential damping that  would be obtained using EDQNM.   
In our case, only small perturbations are damped exponentially. The initial decay of large perturbations is algebraic.

 To see this, let us first follow the optimal decay of an initially large energy perturbation,  say $E_\lo(t) =(1+\delta_\lo(t))/\beta$ with $\delta_\lo(0^+) \gg 1 $ (or alternatively $\beta^0_\lo/\beta \ll 1$). At early times ($t\ll\tau_\kappa(\lo)$), the $\tanh$ decay  follows a power-law dependence :   $\delta_\lo(t) \propto t^{-2} k^{-2}(\lo) $.
  At late times, the damping becomes exponential $\delta_\lo(t) \sim 2 e^{-t/\tau_\kappa(\lo)} $.
  Similarly, any initially small perturbation, say $0 < \delta_\lo(0^+) \ll 1$,  also essentially decays exponentially  :
  $\delta_\lo(t) \sim \delta_\lo(0^+) e^{-t/\tau_\kappa(\lo)}$.  
 In both two regimes, the net effect of the bath could be modeled in terms of 
a  scale-dependent eddy viscosity $\nu_{opt}(\lo)=k^{-2}(\lo) \tau_\kappa^{-1}(\lo)$, which yields the scaling  
$\nu_{opt}(k) \propto k^{-1}\sqrt{E/N}$, where $k$ denotes the magnitude of the $k^{th}$ wavenumber.
In this very simple setting, the optimal closure clearly identifies the relevant time scale as the equilibrium time associated to the bath, without any external modeling. In this sense, the perturbed  energy is dissipated without requiring any artificial eddy damping.

By contrast, were we to tackle the damping from an  EDQNM point of view, the answer would be less definitive. The result depends on the eddy damping introduced to model the transfer of energy between the perturbed shell and its nearest and next-nearest neighbors. 
On the one hand, if the eddy viscosities are built upon  the typical ``equilibrium time'' $\tau_\kappa(\lo)$, we recover the damping rate associated to the small perturbation limit of the optimal framework. 
On the other hand,  we may as well assume that the transfers of energy between the perturbed  shell and the bath are mediated by a constant flux  $\epsilon$, in accordance with the standard phenomenology of turbulence. We then obtain   the scalings $v[k] \propto k^{-1/3}$, and corresponding turnover time $\tau_{\epsilon}[k] \propto k^{-2/3}$.  The eddy viscosity would then scale as $k^{-4/3}$. It is thus hard to decide \emph{a priori} which timescale is the most appropriate in this specific test problem, and hence the EDQNM closure does not produce a clear result. 
The reader interested in the details of the EDQNM closure for the shell dynamics can refer to  \ref{sec:EDQNM}.

\subsection{Thermodynamic interpretation}
We conclude this section with a comment on the ``thermodynamic variable'' $\sigma_\lo$ defined in (\ref{eq:SigmaVar}).  A direct calculation shows that $\sigma_\lo = \avt{\log \tilde \rho} + const. $;  $\sigma_\lo$ can  therefore be interpreted as minus the entropy of the proxy density $\tilde \rho$. Its time derivative is akin to an entropy production rate, which is directly related to the energy dissipation rate of the perturbed mode. In our problem, the energy dissipation rate can be identified as  $\epsilon_\lo = \avt{\mR|v_\lo|^2}$.  In turn, one can check that $\avt{\mR|v_\lo|^2}=-\dot b_\lo/b_\lo^2$,   so that $- \dot \sigma_\lo = b_\lo \epsilon_\lo $: the rate of energy dissipation is the instantaneous temperature times the entropy production. When $t \to \infty$, the energy dissipation rate becomes zero because the system has thermalized,  while the entropy production rate tends to zero because the true ensemble  becomes  Gaussian.

\section{Numerics}
\subsection{Protocol}
In order to test the predictions of the optimal closure, we perform two series of numerical experiments for the shell dynamics  (\ref{eq:CompactShell}), using $N=23$ (set A) and $N=31$ wavenumbers (set B)  with intershell ratio $\lambda=2$, so that $k(n)=k_0 2^n \in [k_{\min};k_{\max}]$. We simulate relaxations towards the energy equipartition ensemble, which we recall is  obtained by setting $\beta=N$ and $\zeta = 0$ in Equation (\ref{eq:gibbs}).  The corresponding  averaged energies and helicities  are then set to  $E_{equi}=1$ and $H_{equi} = (k_0/3)\left( 1-(-2)^N \right) $.  
For each experiment, we randomly generate  ensembles of perturbed initial configurations using the probability distributions $(\ref{eq:Ansatz})$. The perturbed energy  $E_{\lo}(0^+) \in [0 , 1]$ is determined by the initial inverse temperature of the perturbed mode: $b_\lo(0^+) = 1/E_{\lo}(0^+)$. The initial averaged helicity of the perturbed mode is then  $\gamma_\lo E_\lo$.  To guarantee the relaxation towards the equipartition ensemble, the initial temperatures of the $N-1$ other modes need to be chosen so as to compensate the energy and helicity perturbations. A possible choice is to take them as  $b_l(0^+) = b_B + \zeta_B (-1) k(l)$ with $b_B$ and $\zeta_B$ determined by
\begin{equation}
H_{equi}-\gamma_\lo E_\lo = \dfrac{1}{b_B}\sum_{l \neq \lo} \dfrac{\gamma_l}{1 + b_B^{-1}\zeta_B \gamma_l}  \text{~and~} 1-E_{\lo} =\dfrac{1}{b_B}\sum_{l \neq \lo} \dfrac{1}{1 + b_B^{-1}\zeta_B \gamma_l}.
\end{equation}
We  report the results observed for  $E_{\lo} =10^{-1}$, $10^{-2}$ and $10^{-3}$, when perturbing the  wavenumbers $0 \le k(l_0) \le 22$ and averaging over $N_s = 500$ realizations for each ensemble. For low-mode perturbations, the variation of the bath helicity is then very small compared to the total helicity,  $ |\gamma_\lo E_\lo /H_{equi}| < 3 \cdot 2^{\lo-N} \ll 1$. This essentially induces $\zeta_B \simeq 0$, consistent with the approximation $b_l(t) \simeq \beta$ that we used in the previous section.

\begin{figure}
\begin{minipage}{0.49\textwidth}
\centering
\includegraphics[width=\textwidth]{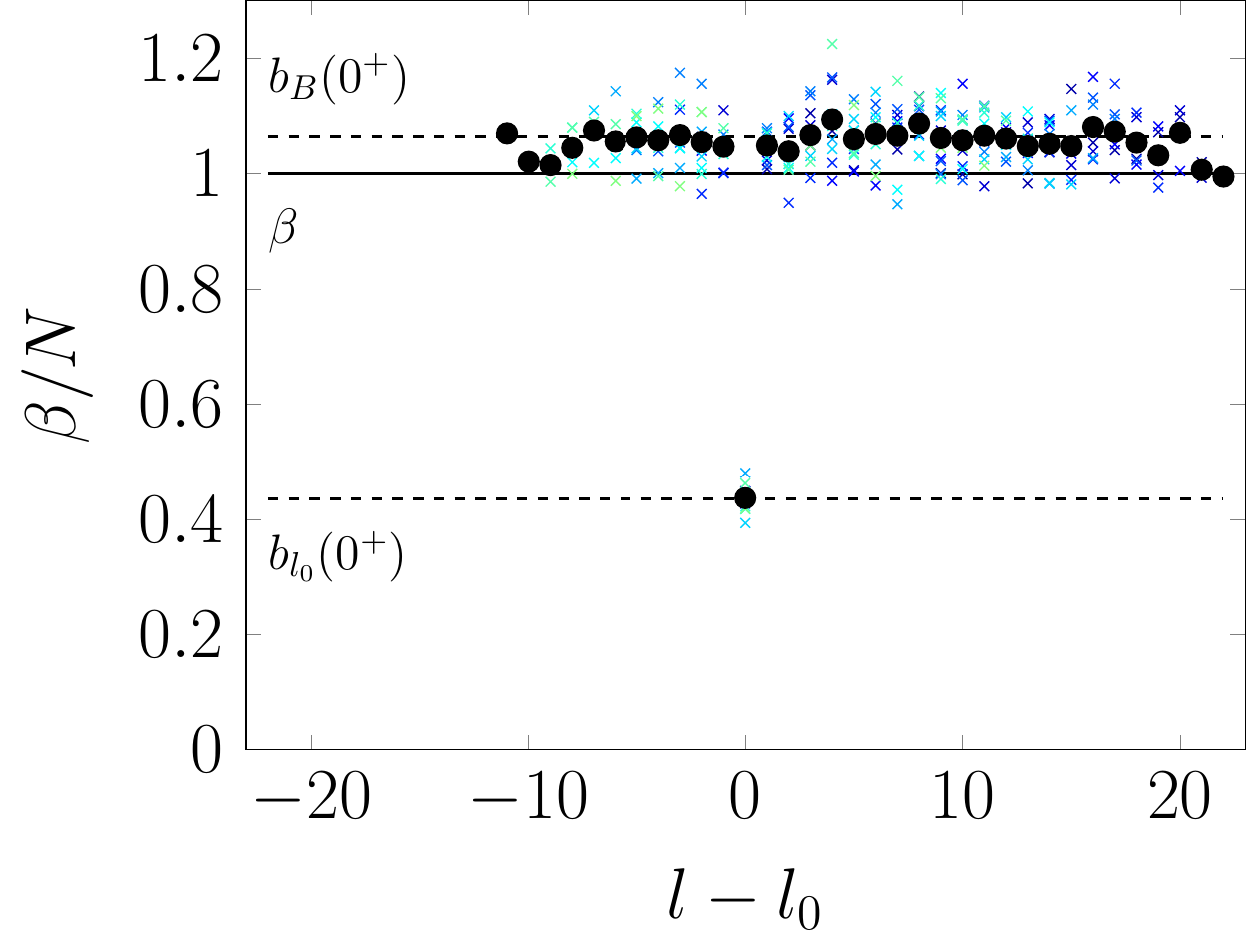}
\end{minipage}
\begin{minipage}{0.49\textwidth}
\centering
\includegraphics[width=\textwidth]{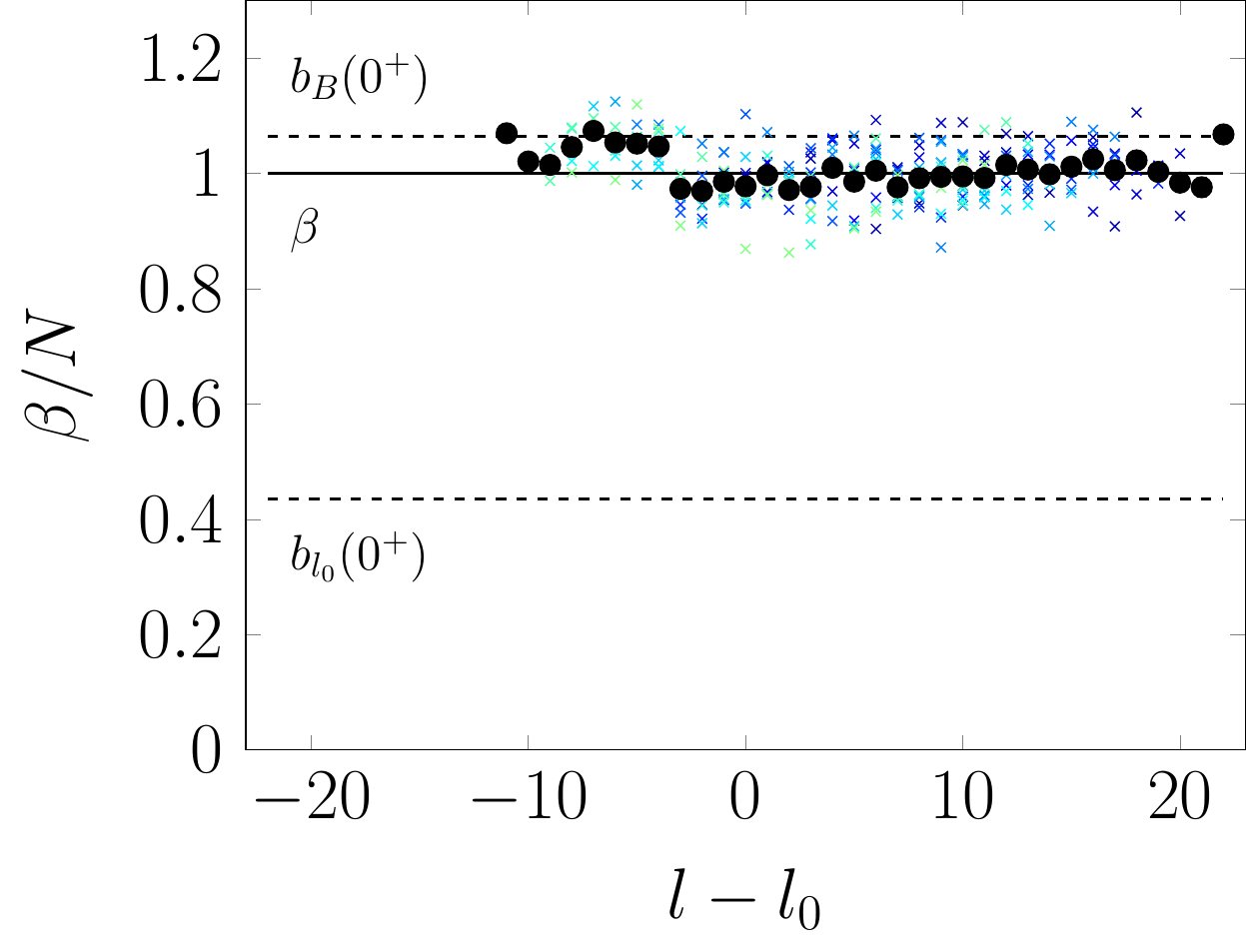}
\end{minipage}
\caption{Inverse temperatures measured at initial times (left) and final times (right) for the set A with $E_{\lo}(0^+)=10^{-1}$. Each set of color represent a different ensemble ( a different perturbed mode). The black dots represent the averages of the inverse temperatures taken over all the ensembles.  It can be observed that at final times  the small-scale modes ($l>l_0$)  have relaxed towards the \modif{equipartition} equilibrium\modif{, represented by the solid black line $\beta=1$}.}
\label{fig:IllusRelax}
\end{figure}

Time integration is performed using a $4^{th}$-order Runge-Kutta scheme, with a constant timestep $\delta t$, that guarantees the conservation of energy and helicity up to $10^{-4}$ during the relaxation for each realization.  With  this numerical protocol,  all the high-frequency modes ($k \ge k(\lo)$) are observed to relax towards their equilibrium values (see Figure \ref{fig:IllusRelax} for an illustration).

\begin{table}
\caption{Parameters used for the numerical experiments.  Each pair $(E_\lo(0^+) ,k(\lo))$  determines an \emph{ensemble} of relaxations, for which $N_s=500$ relaxations are computed. The time of integration for each realization is $T_f/k(l_0)$. }
\label{table:param}
\centering
\begin{tabular}{cccccccccc}
SET & $N$ & $k_0$ & $k_{\min}$ & $ k_{\max}$& $\delta t$ & $E_{equi}/N$ & $\log_2 k(\lo)$ & $E_{\lo}(0^+)$ &$T_f$  \\
\hline 
A & $23$ & $2^{-1}$ &1 & $2^{22}$&$3 \times 10^{-8}$ & $1/23 \simeq 0.044$ & $0 \to 22 $ &$10^{-1},10^{-2},10^{-3}$ & 10\\
B & $31$ & $2^{-8}$ & $2^{-7}$& $2^{22}$& $3 \times 10^{-8}  $ & $1/31 \simeq 0.033$ & $3 \to 22$ & $10^{-1},10^{-2},10^{-3}$ & 5\\
\end{tabular}
\end{table}
\subsection{Results}
\begin{figure}
\begin{minipage}{0.52\textwidth}
\centering
\includegraphics[width=\textwidth]{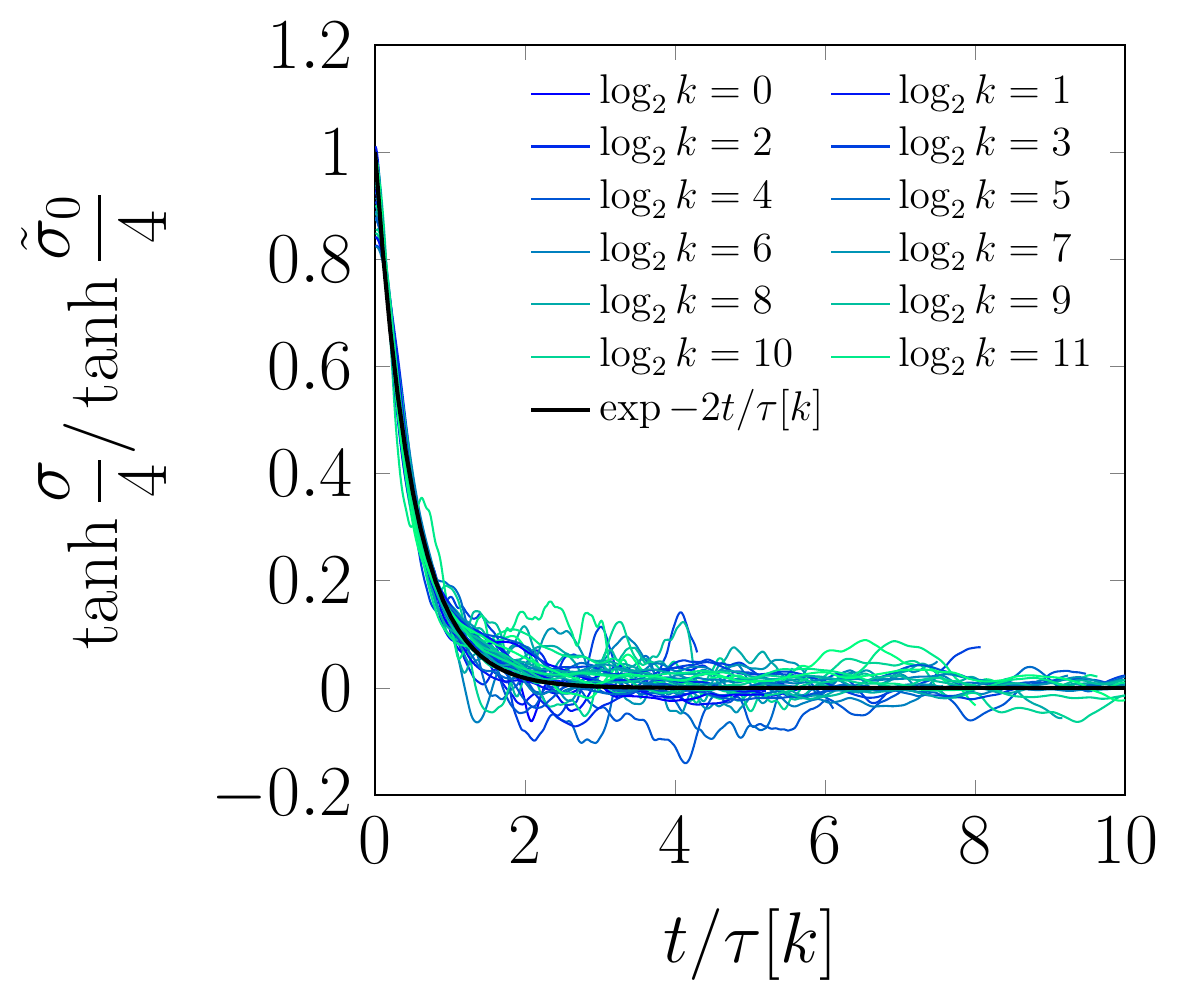}
\end{minipage}
\begin{minipage}{0.48\textwidth}
\centering
\includegraphics[width=\textwidth]{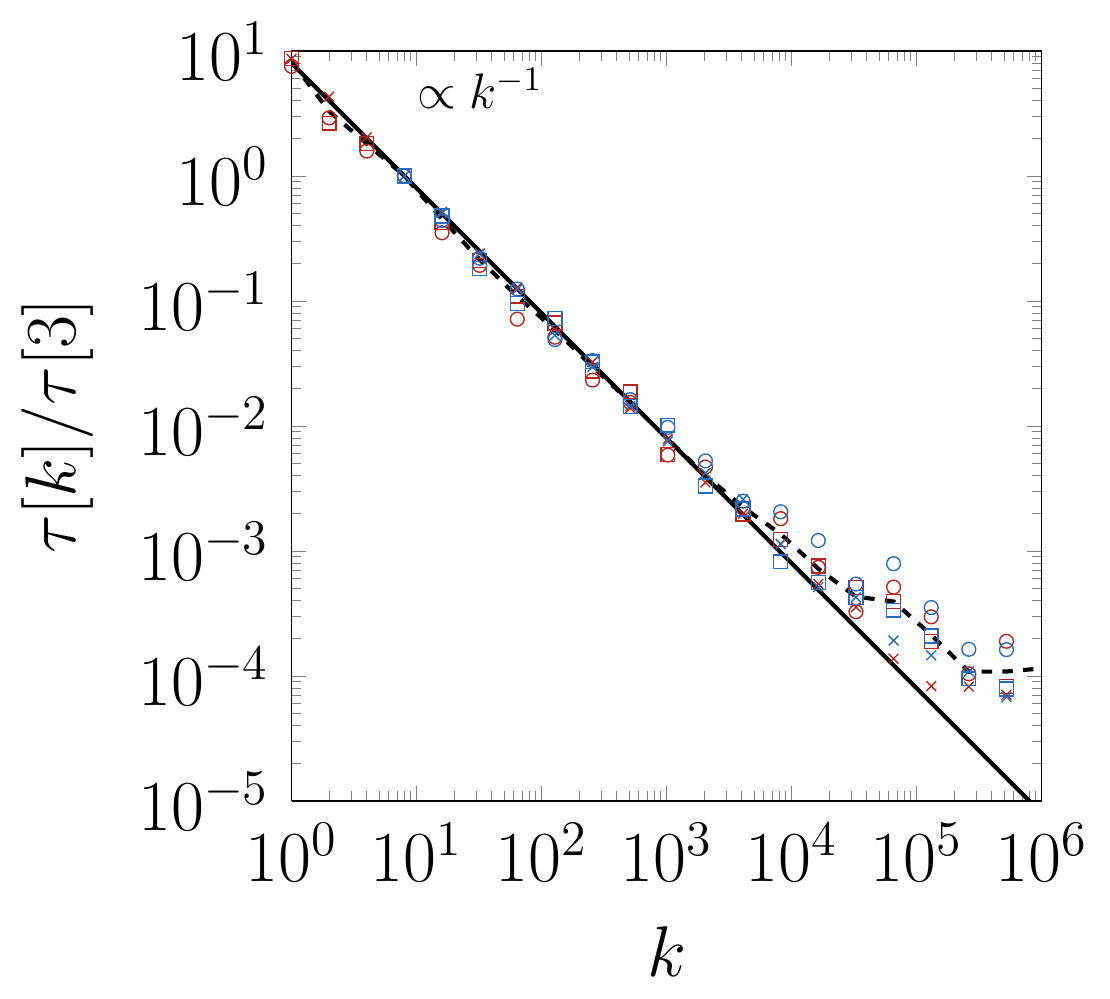}
\end{minipage}
\caption{Left : Relaxation of the entropy $\sigma$ for the set  $A$ and $\log_2 k[l_0] <11$. All the relaxations are here superimposed : to each color correspond three datasets, one for each value of the perturbed energy indicated in Table \ref{table:param} (see text for details). Right : Spectral dependence of the observed relaxation time $\tau[k]$ for sets A (in red) and B (in blue).
The symbols code the initial perturbed energy : $10^{-1}$ ($\circ$), $10^{-2}$ ($\square$), and $10^{-3}$ ($\times$). The dotted line represent the average taken over all the ensembles.}
\label{fig:Collapse}
\end{figure}

The numerics match two main features of the optimal predictions. When the perturbed wavenumbers are low ($k \ll  N$), both the shapes of the relaxation profiles, and their timescales are in good agreement with the predictions of the optimal closure. On Figure \ref{fig:Collapse}, we display the relaxation profiles in terms of the entropy variable $\sigma$, which we recall  is the natural thermodynamic variable for the optimal closure framework.  For each ensemble, we fit  the profile (\ref{eq:relax})  $\tanh \frac{\sigma}{4}  = \tanh \frac{\tilde \sigma_0}{4}  e^{- 2 t/\tau}$ to the averaged DNS profile, by adjusting numerically both $\tilde \sigma_0$ and $\tau$. The numerical data can then be collapsed on the same graph. These data are shown on the left panel of Figure \ref{fig:Collapse}, which gathers together all the relaxations observed for the set A for $k \le 2^{11}$, regardless of the perturbation size. This collapse of the data validates the main quantitative prediction of the optimal closure theory. 

\begin{figure}
\begin{minipage}{0.47\textwidth}
\centering
\includegraphics[width=\textwidth]{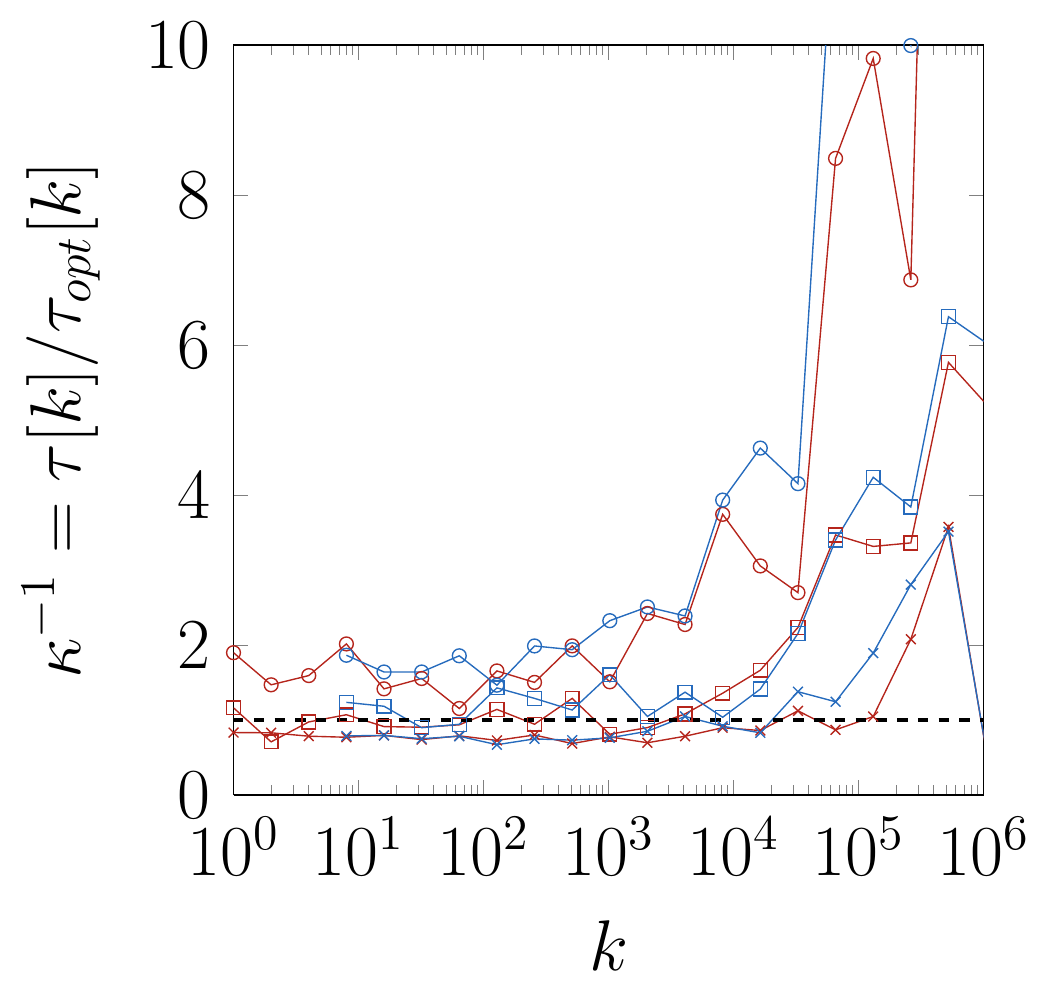}
\end{minipage}
\begin{minipage}{0.525\textwidth}
\centering
\includegraphics[width=\textwidth]{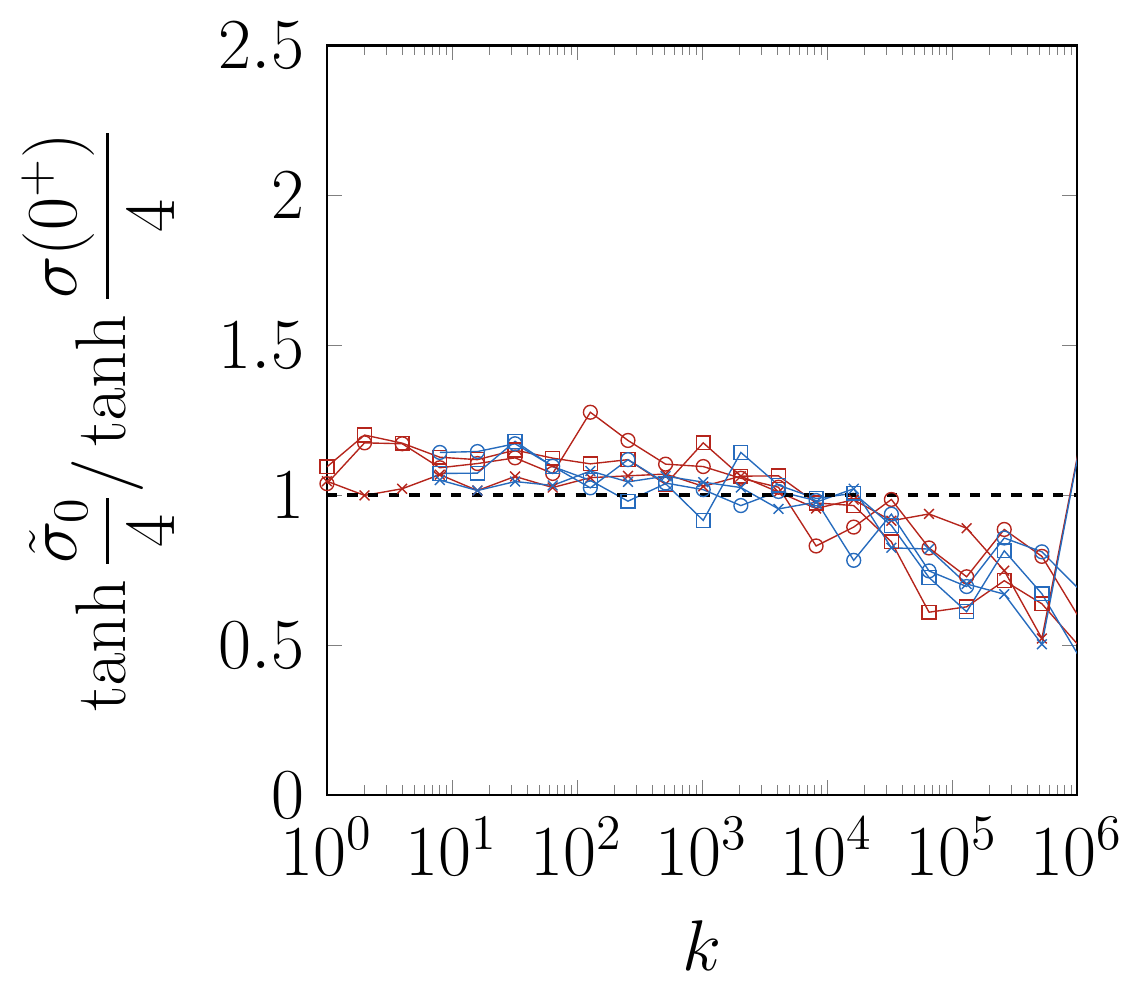}
\end{minipage}
\caption{Left : Spectral dependence of the weighing constant $\kappa$, measured as the ratio between the numerically adjusted time scale and the optimal time obtained with a weighing constant set to 1. Right :  Dependence of the parameter $\tilde \sigma_0$ used for the  collapse shown on Figure  \ref{fig:Collapse}.  
The color code  and the symbols are  the same as for Figure \ref{fig:Collapse}. The symbols code the initial perturbed energy : $10^{-1}$ ($\circ$), $10^{-2}$ ($\square$), and $10^{-3}$ ($\times$). The color codes the set  : red for set A and blue for set  B }
\label{fig:Scaling}
\end{figure}

The right panel of Figure \ref{fig:Collapse} shows  for both sets A and B the spectral dependency of the time scales inferred 
from the numerical experiments. For low to intermediate $k$  (say $k \lessapprox 12 $), they indeed scale as $k^{-1}$, as
anticipated by the optimal formula (\ref{eq:relaxCste}). We note no significant difference between the two sets $A$ and $B$.
For the intermediate scales, a closer inspection reveals that the values found for the parameter $\tilde \sigma_0$  are close to the predicted values, although systematically slightly above, as shown on the right panel of  Figure \ref{fig:Scaling}. The left panel of Figure \ref{fig:Scaling} displays the behavior of the only genuine fitting parameter of the theory, namely the weighing constant $\kappa$ introduced in Equation (\ref{eq:defLOF}). From a practical point of view, the value of  $\kappa$ determines the relaxation time scale, obtained  as $\tau_{\kappa}(k) = 2 (N/69)^{1/2} (\kappa k(\lo) )^{-1}$ from Equations (\ref{eq:relax})  and (\ref{eq:relaxCste}). For each ensemble, we measure $\kappa$ as the ratio between the time $\tau[k]$ previously determined and the optimal time $\tau_{opt}(k) = \tau_{\kappa=1}(\lo) = 2 (N/69)^{1/2} (k(\lo) )^{-1}$. 
For low and  intermediate values of $k$ ($k \lessapprox 12 $), the constant is found to be fairly close to 1 for every ensemble. It does not vary with the resolution $N$, nor does it vary with the shell number, in good agreement with the closure. However, one observes a slight dependency on the perturbed energy not captured by the optimal framework. Averaging $\kappa$ over the first 12 shells, we indeed observe $\kappa = \kappa(E_{\lo})$, to be a decreasing function of the perturbed energy ranging from
 from $\kappa \simeq 1.25$ for $E_{\lo}=10^{-3}$ to $\kappa \simeq 0.5$ for $E_{\lo}=10^{-1}$.
\subsection{Comments}
The numerical results support the predictions of the optimal closure.  This may seem surprising, in view of the crude and simple  Ansatz (\ref{eq:Ansatz}) on which the closure is based.   It could be argued that the test problem set that we have used is designed to generate only small deviations from Gaussianity.     Indeed, we checked that  those deviations develop within one relaxation time, before fading away as the system relaxes to Gibbs equilibrium.  Moreover, no significant deviations from Gaussianity were observed in the statistics of the bath.   Similarly, we monitored the two-mode correlation, which showed only weak correlations  between the perturbed modes and the bath. In our test problem, therefore,  the Ansatz is  not  unsound. 

While the low $k$ behavior show remarkable agreement with the optimal predictions, strong departures are observed for the high modes. We can think of two main reasons.
From a mathematical point of view, perturbation of a  mode with large $k$ will result in a large perturbation in the helicity. 
Hence the approximation $b_l(t) \simeq \beta$ may no longer be valid.  
 More crucially though, the relaxation that the optimal closure intends to describe is caused by a thermalized bath, with the underlying assumption that the latter is composed of fluctuating degrees of freedom.  This assumption does not hold when  the smallest scales are perturbed, as the bath is then only composed of slow degrees of freedom.  Those modes  $l<l_0$  are 
 essentially frozen during the relaxation (see the right panel of Figure \ref{fig:IllusRelax}), and hence we can expect that they
 play no significant role in the thermalization process.  
That the optimal closure fails to capture the relaxation in this situation is not surprising, given that it is suited to the reduction of the  dynamics of the largest scales of the motion only.   %This could also be a reason why the weighing constant $\kappa$ shows a mild dependence on the perturbed energy, that is not captured by the optimal closure.
\section{Perspectives}
This work has exposed a new kind of turbulent closure, which we term ``optimal closure.''  It has been used to predict the relaxation towards a Gibbs thermalized state of an inviscid shell dynamics, which can be thought of as  a minimal model for homogeneous turbulence. 
The optimal closure has two appealing features: it is self-consistent and bears close connections with thermodynamics. Within its framework, the effective dissipation on the large-scales due to unresolved small scales is a consequence of non-vanishing entropy production, which is present even when Gaussians are taken to proxy true ensemble averages.
Numerics shows good agreement with the salient predictions for the single-mode test problem: one being the non-trivial shape of the relaxation profile, the other being the correct time scaling for relaxation.   

While this investigation has focused on one simplified model of turbulence, we think that the optimal theory can be useful more widely.  Most of the calculations made in the derivation of our closure rely on the properties (\ref{eq:CompactShell})-(\ref{eq:ShellCoeffproperties}).  This class of dynamics is in fact very general; it includes, in particular, Burgers equation, as well as 2D and 3D Euler dynamics.   For example, for 2D Galerkin Euler dynamics on any bounded domain, the ``C'' coefficients  would   be  $C_{\bk\bl\bm} = \int \phi_\bk[\phi_\bl,\phi_\bm]$, with $[,]$ the usual Poisson Bracket and $\phi_\bk$'s an orthornormal family of the  Laplacian eigenmodes. 
For 3D Euler dynamics, the helical decomposition of the 3D Euler dynamics considered by  \cite{waleffe_inertial_1993} is also closely analogous to  (\ref{eq:ShellCoeffproperties}), and this  feature has been exploited in the past to consider more realistic shell models of turbulence (\cite{benzi_helical_1996}).
In each of these systems, a Gaussian Ansatz like the one exploited in the present work  will lead to a 
corresponding optimal closure that is almost identical to that derived here.   In fact, the steps leading from  the
Liouville equation (\ref{eq:LiouvilleEq}) to the Lagrangian density (\ref{eq:ShellLagrangian})  will proceed in an entirely
analogous way.

However, one must bear in mind some drawbacks of the optimal closure framework.
In the present paper the optimal closure is analytically tractable, but in more general situations this is not guaranteed.  
 The crux of the issue lies in solving the Euler-Lagrange equations,  or the associated Hamilton-Jacobi equations, for
 the defining optimization problem.   In a companion paper (\cite{turkington_coarse-graining_2015}),  a  closed set of equations for the inhomogeneous statistical dynamics of a set of large-scale 2D Euler modes is derived at the cost of performing a perturbation expansion 
 in the amplitudes of the nonequilibrium disturbances.   Whether the optimal closure can be used efficiently in combination with more complex  Ans\"atze describing far-from-equilibrium and strongly non-Gaussian effects is still an open question, as is the treatment of  viscous effects. These are challenging matters for future work.

\renewcommand{\mp}{{m^\prime}}
\appendix
\section{EDQNM for the shell dynamics}
\label{sec:EDQNM}
To obtain the  EDQNM prediction for the damping of the perturbed mode $\lo$, one first writes  the equation for the second and third order cumulant involved in its dynamics :
 \begin{equation}
\begin{split}
 \dfrac{\d}{\d t�} \av{|v_\lo|^2}  &=2 C_{\lo mn} \gamma_n \Re \left[  \av{v_\lo^\star  v_m^\star  v_n^\star } \right],\\
\dfrac{\d}{\d t�} \av{v_\lo v_m v_n } & =  C_{\lo \mp \np} \gamma_\np \av{v_m v_n  v_\mp^\star  v_\np^\star }  \\
& +  C_{m \mp \np} \gamma_\np \av{v_\lo v_n  v_\mp^\star  v_\np^\star } +  C_{n \mp \np} \gamma_\np \av{v_\lo v_m  v_\mp^\star  v_\np^\star }.
\end{split}
\label{eq:secondmomentbis}
\end{equation}
The EDQNM dynamics is obtained by modifying both sides of the equation for the third moment: the two averages $\av{\cdot}$ and $\avt{\cdot}$ are identified to evaluate the fourth-order moment on the r.h.s and a damping term is added on the l.h.s to compensate for the latter approximation.
The equation that we now use is  
\begin{equation}
\begin{split}
\left[ \dfrac{\d}{\d t�} + \mu_{\lo m n} \right]\av{v_\lo v_m v_n } %& =  C_{\lo \mp \np} \gamma_\np \avt{v_m v_n  v_\mp^\star  v_\np^\star }  \\
%& +  C_{m \mp \np} \gamma_\np \avt{v_\lo v_n  v_\mp^\star  v_\np^\star } +  C_{n \mp \np} \gamma_\np \avt{v_\lo v_m  v_\mp^\star  v_\np^\star }.\\ 
%& = C_{\lo m n} \dfrac{ \gamma_n - \gamma_m}{b_m b_n} + C_{m \lo n}  \dfrac{ \gamma_n - \gamma_\lo}{b_n b_\lo}+ C_{n \lo m}  \dfrac{ \gamma_m - \gamma_\lo}{b_m b_\lo}\\
& = C_{\lo m n} \left[ \dfrac{ \gamma_n - \gamma_m}{b_m b_n} +  \dfrac{ \gamma_\lo - \gamma_n}{b_n b_\lo}+ \dfrac{ \gamma_m - \gamma_\lo}{b_m b_\lo} \right],
\end{split}
\end{equation}
where the coefficient $\mu_{l m n}$ has dimension 1/time. It is meant to capture the damping caused by  by the higher order statistics.
%\begin{equation}
% \mu_{lmn} = \mu_l + \mu_m + \mu_n \text{~with dimension 1/time~}% \text{~and~} \mu_k =1/\tau_\epsilon[k] \propto \epsilon^{1/3} k^{2/3}.
%\end{equation}
%What we take for  the $\mu_k$'s is a crucial but subtle question : we postpone the discussion to the end of the appendix.
Identifying $\avt{\cdot}$ and $\av{\cdot}$ for the l.h.s of the equation for the second moment yields
\begin{equation*}
\begin{split}
  \dfrac{\d}{\d t�} b_\lo^{-1}  &= 2 C_{\lo mn}^2 \gamma_n \int_0^t \d s e^{-\mu_{\lo m n}(t-s)} \left[ \dfrac{ \gamma_n - \gamma_m}{b_m(s) b_n(s)} +  \dfrac{ \gamma_\lo - \gamma_n}{b_n(s) b_\lo(s)}+ \dfrac{ \gamma_m - \gamma_\lo}{b_m(s) b_\lo(s)} \right],\\
& \simeq 2 C_{\lo mn}^2 \gamma_n \dfrac{1}{\mu_{\lo m n}} \left[ \dfrac{ \gamma_n - \gamma_m}{b_m(t) b_n(t)} +  \dfrac{ \gamma_\lo - \gamma_n}{b_n(t) b_\lo(t)}+ \dfrac{ \gamma_m - \gamma_\lo}{b_m(t) b_\lo(t)} \right] , \\
\end{split}
\end{equation*}
where the Markovian approximation is implemented to make the final approximation.   

Further assuming $b_m(t) =�\beta$ for $l \neq \lo$, we get the evolution of the perturbed energy in  terms of the the function $f$ previously defined  by (\ref{eq:fg}) :
%\begin{equation}
%  \dfrac{\d}{\d t�} b_\lo^{-1}  \simeq 2 C_{\lo mn}^2 \gamma_n \dfrac{e^{-\mu_{\lo m n }t}-1}{\mu_{\lo m n}} \left[ \dfrac{ \gamma_n - \gamma_m}%{\beta_{equi}^2} +  \dfrac{ \gamma_\lo - \gamma_n}{\beta_{equi} b_\lo(t)}+ \dfrac{ \gamma_m - \gamma_\lo}{\beta_{equi} b_\lo(t)} \right] 
%\end{equation}
%
%
\begin{equation}
  \dfrac{\d}{\d t�} b_\lo^{-1}  \simeq \dfrac{2}{\beta}\left[ \dfrac{ 1}{\beta} -  \dfrac{1}{b_\lo(t)}\right] \sum_{1 \le m,n \le N} \dfrac{f[\lo,m,n]}{\mu_{\lo m n}} . 
\end{equation}
%with $f$ the function previously defined  by (\ref{eq:fg}).
%
%
If the $\mu_{\lo m n}$ do not depend on the perturbed energy, then the evolution of the energy perturbation $\delta_\lo(t) = \beta/ b_\lo(t) -1$ is exponential : 
\begin{equation}
\delta_\lo(t) = \delta_\lo(0^+) e^{ -2 t/\tau_{edqnm}(\lo)} \text{~with~}  \tau_{edqnm}(\lo) = \beta \left(\sum_{1 \le m,n \le N} \dfrac{f[\lo,m,n]}{\mu_{\lo m n}} \right)^{-1}.
\end{equation}
The effective dissipation rate depends on what we choose to be the damping times $1/\mu_{\lo mn}$ associated to triads of neighboring shells $(\lo,m,n)$. A standard choice would express the $\mu_{lmn}'s$   as
\begin{equation}
 \mu_{lmn} = \mu_l + \mu_m + \mu_n,
\end{equation}
with the $\mu_l$ representing the typical frequency associated  to the shell $l$.
In our case, it is however not self-evident  what those are.
On the one hand, we may assume that the typical time is an equilibrium time so that $\mu(l) \sim k(l) b(l)^{-1/2}$ (up to a non-dimensional constant). For  \emph{small perturbations}, this yields $\mu(l) \sim  k(l)  E^{1/2}N^{-1/2}$, and hence $\tau_{edqnm}(\lo) \sim N^{1/2}E^{-1/2} k(\lo) \phi_{\kappa=1}(\lo)^{-1}$. Up to a non-dimensional constant, the expression is the same as (\ref{eq:relax}), and  the scaling is $\tau_{edqnm}(k) \sim N^{1/2}E^{-1/2} k^{-1}$. 
On the other hand, we may as well assume that the transfers between the perturbed mode and its neighboring shells is given by a Kolmogorov phenomenology. The asssumption of  a constant rate of energy transfer $\epsilon$ yields a scaling $v(l) \sim \epsilon^{1/3} k(l)^{-1/3}$ and therefore a typical frequency $\mu(l) \sim  \epsilon^{1/3} k(l)^{2/3}$. The effective damping time would be  $\tau_{edqnm}(\lo) \sim NE^{-1} \epsilon^{1/3} k(\lo)^{2/3}\phi_{\kappa=1}(\lo)^{-1}$, with scaling $\tau_{edqnm}(k) \sim NE^{-1}\epsilon^{1/3} k^{-4/3}$.
 
 \section*{Acknowledgments}
The work reported in this paper was partially supported by the National Science Foundation under grant DMS-1312576.

\section*{References}
\bibliographystyle{iopart-num}
\bibliography{NotesOnOptimalRelax}
\end{document}